\documentclass[prl,aps,twocolumn,showpacs,superscriptaddress]{revtex4-1}
\usepackage{hyperref}
\hypersetup{colorlinks=true, citecolor=blue, filecolor= black, linkcolor= blue, urlcolor= blue}
\usepackage{graphicx}
\usepackage{dcolumn}
\usepackage{bm}
\begin{document}


\title{Optical Forces of Focused Ultrafast Laser Pulses on Nonlinear Optical Rayleigh Particles}

\author{Liping Gong}

\author{Bing Gu}
\email{gubing@seu.edu.cn}

\author{Guanghao Rui}

\author{Yiping Cui}

\affiliation{Advanced Photonics Center, Southeast University, Nanjing 210096, Jiangsu, China }

\author{Zhuqing Zhu}
\affiliation{Key Laboratory of Optoelectronic Technology of Jiangsu Province, School of Physical Science and Technology, Nanjing Normal University, Nanjing 210023, Jiangsu, China }

\author{Qiwen Zhan}
\affiliation{Department of Electro-Optics and Photonics, University of Dayton, 300 College Park, Dayton, OH, 45469-2951 USA }


\begin{abstract}
\noindent The principle of optical trapping is conventionally based on the interaction of optical fields with linear induced polarizations. However, the optical force originating from the nonlinear polarization becomes significant when nonlinear optical nanoparticles are trapped by ultrafast laser pulses. Herein we establish the time-averaged optical forces on a nonlinear optical nanoparticle using high-repetition-rate ultrafast laser pulses, based on the linear and nonlinear polarization effects. We investigate the dependence of the optical forces on the magnitudes and signs of the refractive nonlinearities. It is found that the self-focusing effect enhances the trapping ability, whereas the self-defocusing effect leads to the splitting of potential well at the focal plane and destabilizes the optical trap. Our results show good agreement with the reported experimental observations and provide a theoretical support for capturing nonlinear optical particles.
\end{abstract}



\maketitle
Optical trapping, also known as optical tweezers, is a useful technique for noncontact and noninvasive manipulation of small particles using a focused laser beam \cite{R01}. This technique has wide applications in physics, chemistry, biology, and other disciplines \cite{R02, R03}. Up to now, the stable optical trapping of micro- and nanoparticles have been extensively demonstrated by the use of continuous-wave (CW) Gaussian laser beam \cite{R01, R02, R03}, cylindrical vector beam \cite{R04}, evanescent field \cite{R05}, plasmonic field \cite{R06}, spinning light fields \cite{R07}, etc. Lots of efforts have been devoted to trap a variety of small objects, such as dielectric particles \cite{R01}, metallic Rayleigh nanoparticles \cite{R04}, semiconductor quantum dots \cite{R08}, and biological cells \cite{R09}.

Recently, optical trapping technique has been extended by substituting CW laser with high-repetition-rate ultrafast laser pulses \cite{R10, R11, R12}. With the ultrafast laser pulses, several novel phenomena have been observed, including the trapping split behavior in the process of capturing gold nanoparticles by ultrafast near-infrared laser pulses \cite{R13}, a controllable directional ejection of optically trapped nanoparticles \cite{R14}, and the immobilization dynamics of a single polystyrene sphere \cite{R15}. It should be noted that the optical force originating from the nonlinear polarization becomes significant and cannot be neglected if the trapped particles exhibit nonlinear optical effects. Moreover, the experimental observations have revealed that the nonlinear optical effects could enhance the optical force \cite{R08, R16} or modify the optical trapping potential \cite{R13}.

To quantitatively appraise the trapping ability, the optical force exerted on a spherical nanoparticle arising from the linear polarization has been calculated using various approaches, such as Rayleigh scattering formulae \cite{R01}, Maxwell's stress tensor \cite{R17}, and discrete dipole approximation \cite{R18}. For a nonlinear optical Rayleigh particle, however, the optical force unambiguously originates from the contribution of both the linear and nonlinear induced polarizations. To the best of our knowledge, there is no theoretical report on the optical force exerted on a nonlinear optical Rayleigh particle beyond the linear optical regime. Alternatively, researchers directly modified the Rayleigh scattering formulae by replacing the refractive index $n_0^p$ with $n_0^p + n_2 I$ \cite{R19, R20, R21}, where $I$ is the optical intensity, $n_0^p$ and $n_2$ are the linear and third-order nonlinear refractive indexes of particle, respectively. This phenomenological theory could interpret the self-focusing effect that increases the trapping force strength and improves the confinement of Rayleigh particles \cite{R19, R20, R21}.

In this work, for the first time, we establish the time-averaged optical forces on a nonlinear optical Rayleigh particle using high-repetition-rate ultrafast laser pulses, based on the linear and nonlinear polarization effects.

For time-harmonic electromagnetic waves with a Gaussian temporal envelope, we have

\noindent
\begin{equation} \label{eq1}
\vec{E} (\vec{r},t) = \vec{E}_0 (\vec{r})\exp (-i \omega t) \exp [-2 \ln (2) t^2 /\tau_F^2],
\end{equation}

\noindent
\begin{equation} \label{eq2}
\vec{B} (\vec{r},t) = \frac {c}{i \omega} \nabla \times \vec{E} (\vec{r},t),
\end{equation}

\noindent where $\vec{E}_0 (\vec{r})$ is the complex function of position in space, $\omega$ is the circular frequency, $c$ is the speed of light in vacuum, and $\tau_F$ is the full width at half maximum for a Gaussian pulse.

Now we consider a spherical dielectric particle immersed in liquid (e.g. water) exhibiting a linear susceptibility $\chi_1$ and a third-order nonlinear optical susceptibility $\chi_3$. Besides, we assume that the optical nonlinearity instantaneously responds to laser pulses. According to the Claussius-Mossotti equation, we obtain the particle-induced dipole moment originating from both the linear and nonlinear polarizations as

\noindent
\begin{equation} \label{eq3}
\vec{p} (\vec{r},t) = 4 \pi a^3 \varepsilon_0 \frac {(\chi_1 + \chi_3 |\vec{E} (\vec{r},t)|^2 ) \vec{E} (\vec{r},t)}{\chi_1 + \chi_3 |\vec{E} (\vec{r},t)|^2+3},
\end{equation}

\noindent where $a$ is the radius of the particle and $\varepsilon_0$ is the permittivity of free space.

Under the excitation of ultrafast laser pulses, the instantaneous optical force exerting on the Rayleigh particle ($a\ll \lambda$) for which time-averaging over one pulse period $T$ yields \cite{R22}

\begin{eqnarray}\label{eq4}
\nonumber \langle \vec{F} \rangle = & \frac {1}{4 T} \int_{-T/2}^{T/2} [(\vec{p} + \vec{p}^*)\cdot \nabla (\vec{E} + \vec{E}^*)  \\
 &  + \frac {1}{c} {\left ( \frac {\partial \vec{p}}{\partial t} + \frac {\partial \vec{p}*}{\partial t} \right )} \times (\vec{B} +\vec{B}^*)     ] dt,
\end{eqnarray}

\noindent where $^*$ denotes the complex conjugate.

The stable optical trapping of nanoparticles with high-repetition-rate ultrafast laser pulses have been experimentally demonstrated \cite{R10, R12}. Each ultrafast laser pulse leads to instantaneous trapping of a nanoparticle. The high-repetition-rate ensures repetitive trapping by successive pulses, and hence the particle does not diffuse significantly between pulses. Typically, the pulse duration $\tau_F$ and repetition-rate $\nu$ (i.e., the inverse of the pulse period $T$) for a commercial Ti:sapphire oscillator are $\sim 100$ fs and 76 MHz, respectively. Substituting Eqs. (1)-(3) into Eq. (4) and considering the condition of $1/(\tau_F \nu)\rightarrow \infty $, we have

\noindent
\begin{equation} \label{eq5}
\langle \vec{F} \rangle = \frac {\pi \varepsilon_0 a^3 \tau_{F} \nu}{\sqrt{\ln 2}} {\rm Re} {\left [ \beta (\vec{E}_0 \cdot \nabla \vec{E}_0^* + \vec{E}_0 \times \nabla \times \vec{E}_0^*) \right ]},
\end{equation}

\noindent with

\noindent
\begin{equation} \label{eq6}
\beta = \int_{-\infty}^\infty \frac {\chi_1 + \chi_3 |\vec{E}_0|^2 \exp (-t'^2)}{\chi_1 + \chi_3 |\vec{E}_0|^2 \exp (-t'^2)+3} \exp (-t'^2) dt'.
\end{equation}

\noindent Here $\chi_1 = \varepsilon_2^0 /\varepsilon_1^0 -1$, where $\varepsilon_2^0$ and $\varepsilon_1^0$ are the permittivities of the particle and the surrounding medium, respectively.

After integrating Eq. (6), we obtain the time-averaged optical forces on a nonlinear optical Rayleigh particle as

\noindent
\begin{equation} \label{eq7}
\langle \vec{F} \rangle  = \frac {1}{4} {\rm Re} (\alpha) \nabla |\vec{E}_0|^2 + \frac {k}{\varepsilon_0 c} {\rm Im} (\alpha) \langle \vec{S} \rangle_{\rm Orb}
\end{equation}

\noindent where $k = 2 \pi / \lambda$ is the wavenumber, $\lambda$ is the wavelength and

\noindent
\begin{equation} \label{eq8}
\langle \vec{S}\rangle _{\rm Orb}  = \langle \vec{S} \rangle +\frac {\varepsilon_0 c}{2 k} {\rm Im} [(\vec{E}_0^* \cdot \nabla)\vec{E}_0],
\end{equation}

\noindent
\begin{equation} \label{eq9}
\langle \vec{S}\rangle  = \frac {1}{2 \mu_0 \omega} {\rm Im} [\vec{E}_0 \times (\nabla \times \vec{E}_0^* ) ],
\end{equation}

\noindent
\begin{equation} \label{eq10}
\alpha  =  \frac {\sqrt{\pi} \tau_F \nu}{2 \sqrt{ \ln 2}} \cdot 4 \pi \varepsilon_0 a^3 \gamma,
\end{equation}

\noindent
\begin{equation} \label{eq11}
\gamma  =  \frac{\varepsilon_2^0 /\varepsilon_1^0 -1}{\varepsilon_2^0 /\varepsilon_1^0 +2} + 3 \sum_{m=2}^\infty \frac {(-1)^m}{\sqrt{m}} \frac {(\chi_3 |\vec{E}_0|^2)^{m-1}}{(\varepsilon_2^0 /\varepsilon_1^0 +2)^m}.
\end{equation}

As described by Eq. (7), optical forces on a nonlinear nanoparticle with ultrafast laser pulses are divided into two parts: the gradient force which is proportional to the gradient of intensity and drives the particle toward the equilibrium point, and the radiation force which is proportional to the orbital part of the Poynting vector of the field and destabilizes the trap by pushing the particle away from the focal point. Note that Eq. (7) degenerates into the one reported previously \cite{R17} for a Rayleigh particle without optical nonlinearity (i.e. $\chi_3 =0$). Moreover, it is demonstrated, both experimentally \cite{R10} and theoretically \cite{R23, R24}, that optical forces due to pulsed beam and CW beam of the same average power should be identical if the nonlinear optical effect can be neglected. Different from the conventionally optical forces arising from the interaction of optical fields with the linear polarization, the optical trapping of nonlinear optical particles originates from both the linear and nonlinear polarizations. Furthermore, the optical forces exerted on the nonlinear particle strongly depend on the nonlinear susceptibility of particle and the distribution of electric field besides the other parameters related to the laser characteristics and the particle itself.

To trap and manipulate nanoparticles in optical tweezers, generally, an $x$-polarized Gaussian beam is tightly focused by a high numerical-aperture (NA) objective lens \cite{R13}. Theoretically, the electric field distribution $\vec{E}_0 (\vec{r})$ in the focal region of an aplanatic lens can be obtained \cite{R25}. Substituting $\vec{E}_0 (\vec{r})$ into Eq. (7), one could calculate the time-averaged optical forces on an optical nonlinear Rayleigh particle by tightly focused linearly polarized Gaussian beam. It should be noted that the value of $\varepsilon_0 c {\rm Im }[(\vec{E}_0^* \cdot \nabla)\vec{E}_0]/(2k)$ in Eq. (8) is proportional to the nonuniform distribution of the spin density of the light field \cite{R17}. When the input beam is a linearly polarized beam, however, this value becomes negligible and can be ignored. Hence, the radiation force expressed by Eq. (7) is only proportional to the Poynting vector $\langle \vec{S} \rangle$ in the following analysis.

In addition to the characteristics of the light field, the polarization induced by the external optical field in the particle undoubtedly plays a crucial role in the magnitude and distribution of the optical force. However, as described by Eq. (11), the nonlinear polarization is related to both the third-order nonlinear optical susceptibility $\chi_3$ and the distribution of electric field. For the sake of simplicity, we only consider Rayleigh dielectric particle exhibiting the Kerr nonlinearity. In this case, the third-order nonlinear optical susceptibility $\chi_3$ is related to the third-order nonlinear refractive index $n_2$ through the conversion formula ${\rm Re } [\chi_3] = n_2 \varepsilon_2^0 \varepsilon_0 c /(3 \varepsilon_1^0 )$. In this work, the nonlinear refractive indexes are taken to be $n_2 = 5.9 \times 10^{-17}$ and $-5.9 \times 10^{-17}$ m$^2$/W, which corresponds to the self-focusing and self-defocusing effects of the trapped particle, respectively \cite{R21}. Without loss of generality, the particle is assumed to be immersed in water. Consequently, the linear refractive indexes of water and particle are taken to be $n_0^w = \sqrt{\varepsilon_1^0}  =1.33$ and $n_0^p = \sqrt{\varepsilon_2^0} = 1.58 +0.01 i$, respectively. For numerical calculations, the other parameters are chosen to be $\lambda = 800$ nm, $NA=0.85$, $\tau_F = 100$ fs, $\nu = 76$ MHz, $a=40$ nm, and the average-power of laser pulses $P=100$ mW in the entire analysis. Using Eq. (7), we investigated both the distribution and the magnitude of the optical forces exerted on the nonlinear optical Rayleigh particle by tuning the sign and magnitude of the nonlinear refractive indexes.

\begin{figure}[h]
\begin{center}
\includegraphics[width=8cm]{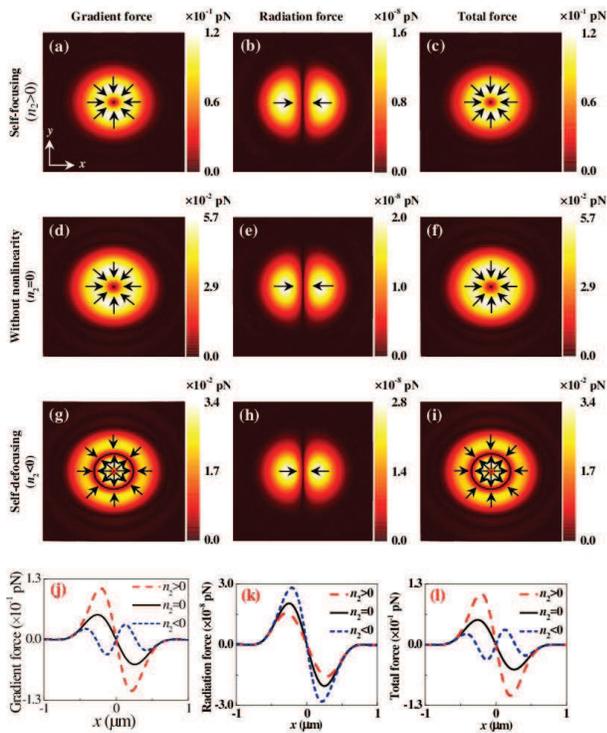}
\caption{%
(Color online) Transverse force distributions produced by tightly focused laser pulses for the particle with self-focusing (first row), without nonlinearity
(second row), and with self-defocusing (third row) in the $x-y$ plane ($z=0$). The bottom row gives the force profiles along the $x$ direction shown in the
above three rows. The arrows in the figures denote the directions of the transverse forces.}
\end{center}
\end{figure}

Figure 1 shows the distributions of the transverse forces produced by tightly focused laser pulses for the particle with self-focusing and self-defocusing effects in the $x-y$ plane ($z=0$). For comparison, both the magnitude and the distribution of the optical force on the particle without nonlinearity ($n_2 =0$) are also shown in the second row of Fig. 1. It is shown that the distributions of the gradient force nearly maintain the circular symmetry in the transverse plane, whereas the corresponding radiation forces exhibit a fan-shaped structure with the two-fold rotation symmetry. It is noteworthy that the magnitude of the radiation forces is negligible compared with that of the gradient forces. Consequently, the distribution of total forces can be nearly regarded to be circularly.

For the particle with self-focusing effect (i.e., $n_2 >0$), the distribution of forces displayed in the first row of Fig. 1 is almost identical to that of the particle without nonlinearity ($n_2 =0$) shown in the second row of Fig. 1. Moreover, the total force always directs to the position of the focal point to produce a force balance. However, the magnitude of the total force exerted on the self-focusing particle is about two times larger than that of the linear particle. Obviously, the nonlinear polarization arising from the self-focusing effect improves the particle trapping ability, which is consistent with the reported experimental observations \cite{R08, R16}. Under the excitation of intense ultrafast laser pulses, the effective refractive index $n_0^p + n_2 I(r,\varphi,z)$ of the self-focusing particle ($n_2 >0$) is larger than that of the linear particle ($n_2 =0$). Accordingly, the self-focusing particle bends stronger the light than that of the same particle without nonlinearity under the same illumination condition, resulting in stronger gradient force exerted on the self-focusing particle than that of the linear particle \cite{R26}.

For the self-defocusing particle (i.e., $n_2 <0$), interestingly, both the distribution and the magnitude of the transverse force are quietly different from those of the self-focusing particles and the linear particle. As shown in Fig. 1(i), the distribution of the total force exerted on the self-defocusing particle exhibits the double-ring pattern. Furthermore, the magnitude of the total force on the self-defocusing particle is smaller than that on the linear particle ($n_2 =0$). This special distribution of the transverse force shown in Fig. 1(i) can be interpreted for the following reasons. For the case of $n_2 <0$, a spatial-variant Gaussian-intensity distribution in transverse can induce a negative lens effect due to the presence of space-dependent refractive-index change $n_2 I(r,\varphi,z)$. Hence, the contributions of the linear and nonlinear polarizations to the gradient force are opposite to each other. At relative weak intensity, the nonlinear polarization is not large enough and the force exerted on the nonlinear particle is similar to that of the linear particle except for the reduced force. Under the excitation of high intensity, however, the force direction at the center and edge of the focused beam will be reversed, giving rise to the double-ring distribution of the transverse force.

\begin{figure}[h]
\begin{center}
\includegraphics[width=8cm]{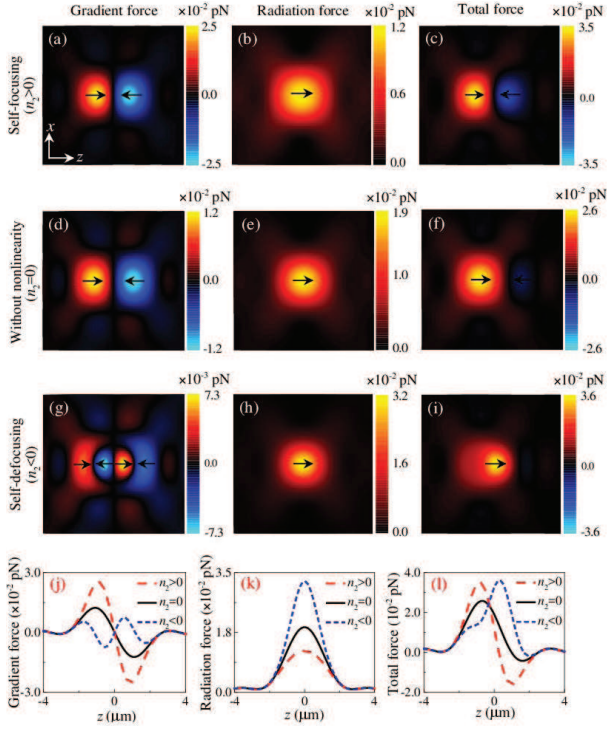}
\caption{%
(Color online) Longitudinal force distributions produced by tightly focused laser pulses for the particle with self-focusing (first row), without nonlinearity
(second row), and with self-defocusing (third row) in the $x-z$ plane ($y=0$). The bottom row gives the force profiles along the z direction shown in the above
three rows. The arrows in the figures denote the directions of the longitudinal forces.}
\end{center}
\end{figure}

Figure 2 illustrates the distributions of the longitudinal forces on the nonlinear particles produced by focused laser pulses in the $x-z$ plane ($y=0$). Different from the transverse forces in the focal plane mainly originating from the gradient forces shown in Fig. 1, the magnitude of longitudinal gradient forces is comparable to those of the radiation forces. Owing to the self-focusing effect of the particle, the longitudinal gradient force increases and the radiation force decreases, contrast to those of the linear particle. As a result, the self-focusing particle with ultrafast laser pulses forms a stable three-dimensional trap in the focal region, which has been validated by the reported experiment \cite{R27}. On the contrary, for the particle exhibiting the self-defocusing effect, both the decreased longitudinal gradient force and the enhanced radiation force destabilize the trap by pushing the particle away from the focal plane. That is, at relatively low power of the laser pulses, the self-defocusing particle is trapped at the focal plane because of the weak optical nonlinearity. With increasing the power under the pulsed excitation, interestingly, the trapped particle will be ejected axially along the direction of the beam's propagation owing to the strong longitudinal force arising from the self-defocusing effect. This theoretical predication successfully explains the puzzling experimental phenomena that the optical trapping is destabilized with increasing power under pulsed excitation only \cite{R12, R15}.

\begin{figure}[h]
\begin{center}
\includegraphics[width=8cm]{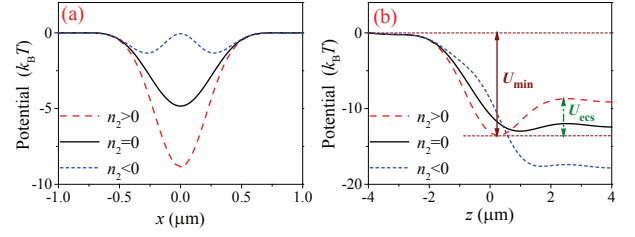}
\caption{%
(Color online) Trapping potential along (a) $x$ direction and (b) $z$ direction with different values of $n_2$.}
\end{center}
\end{figure}

To form a stable particle trap, the potential generated by the gradient force must be deep enough to overcome the kinetic energy of the particle in Brownian motion, $k_B T$, where $k_B$ is Boltzmann constant and $T$ is the absolute temperature of the ambience. Figure 3 shows the trapping potential along the $x-$ and $z-$directions for the particles with different values of $n_2$. As shown in Fig. 3(a), the potential depths for three types of particles are larger than 1 $k_B T$. Accordingly, the particles with and without nonlinearity can be captured along $x$ direction. The potential of the self-focusing particle becomes deeper than that of the linear particle, indicating that the self-focusing effect enhances the trapping ability. Surprisingly, for the particle exhibiting the self-defocusing nonlinearity, the potential splits into two off-axis trapping sites along the x direction, which is supported by the reported experiments \cite{R13}. For the trapping potential along the $z$ direction shown in Fig. 3(b), there are two important parameters: one is the absolute depth of the potential minimum $U_{\rm min}$, another is the potential barrier $U_{\rm esc}$ that directly relates to the trapping efficiency. As shown in Fig. 3(b), the self-focusing effect increases the values of $U_{\rm min}$ and $U_{\rm esc}$ and then improves the confinement of particles. However, the potential barrier $U_{\rm esc}$ of the self-defocusing particle is less than 1 $k_B T$, resulting in the destabilization of the particle in the axial direction.

In summary, for the first time, we have established the time-averaged optical forces exerted on a nonlinear optical Rayleigh particle using high-repetition-rate ultrafast laser pulses, based on the linear and nonlinear polarizations. We have investigated the characteristics of the transverse and longitudinal optical forces for particles exhibiting self-focusing and self-defocusing effects. It is shown that the self-focusing effect increases the trapping force strength and improves the confinement of particles, whereas the self-defocusing effect leads to the splitting of potential well at the focal plane and destabilizes the optical trap resulting in ejections of trapped particles along the direction of the beam's propagation. Our results successfully explain the reported experimental observations and provide a theoretical support for capturing nonlinear nanoparticles with femtosecond laser trapping.

This work was financially supported by National Science Foundation of China (Grant Nos: 11774055, 11474052, 11504049, 61535003), Natural Science Foundation of Jiangsu Province (BK20171364), and National Key Basic Research Program of China (2015CB352002).



\end{document}